\documentclass[journal]{IEEEtran}
\ifCLASSINFOpdf
  \usepackage[pdftex]{graphicx}
\else
  \usepackage[dvips]{graphicx}
\fi
\usepackage[cmex10]{amsmath}
\usepackage[font=footnotesize]{subfig}
\usepackage{stfloats}
\begin{document}
%
\title{MIMO Physical Layer Network Coding Based on VBLAST Detection}
%
%
%

\author{Shengli~Zhang,~
        Canping~Nie,~
        Liya~Lu,
        and~Gongbin~Qian,~
\thanks{S. Zhang C. Nie, L. Lu and G. Qian are with the Shenzhen MCSP Key Lab, Department
of Communication Engineering, Shenzhen University, Shenzhen, China. e-mail:(zsl)@szu.edu.cn.}
\thanks{Manuscript received April 19, 2005; revised January 11, 2007.}}

%
%

\markboth{submitted to IEEE wireless communications lettes }%
{Shell \MakeLowercase{\textit{et al.}}: Bare Demo of IEEEtran.cls for Journals}
%



\maketitle

\begin{abstract}
For MIMO two-way relay channel, this paper proposes a novel scheme, VBLAST-PNC, to transform the two superimposed packets received by the relay to their network coding form. Different from traditional schemes, which tries to detect each packet before network coding them, VBLAST-PNC detects the summation of the two packets before network coding. In particular, after firstly detecting the second layer signal in 2-by-2 MIMO system with VBLAST, we only cancel part of the detected signal, rather than canceling all the components, from the first layer. Then we directly map the obtained signal, summation of the first layer and the second layer, to their network coding form. With such partial interference cancellation, the error propagation effect is mitigated and the performance is thus improved as shown in simulations.
\end{abstract}

\begin{IEEEkeywords}
Multiple Input Multiple Output, Physical layer Network Coding, Two Way Relay Channel, VBLAST.
\end{IEEEkeywords}

%
\IEEEpeerreviewmaketitle

\section{Introduction}
%
%
%
%
\IEEEPARstart{I}{n}
wireless networks, the use of relay is attracting increasing attention
[1, 2] because of its many advantages. Among the relay channels, two-way
relay channel (TWRC, as shown in Fig. 1) is a especially interested due to the almost double spectral
efficiency with the physical layer network coding (PNC) [3] transmission scheme. It was
further proved in [4, 5] that PNC can approach the capacity of TWRC in high
SNR region.

Another spectral efficiency boosting technique is Multiple Input and
Multiple Output (MIMO), which has been widely used in wireless systems.
Therefore, it is of great interest to combine PNC and MIMO to further
improve the wireless spectral efficiency. A straightforward way is to divide
the MIMO transmission into parallel SISO streams by precoding, so that PNC
can be implemented on each steam [6]. However, the precoding requires not
only transmitter side channel state information (CSIT) but also strict time
and carrier phase synchronization between the two end nodes.

MIMO NC scheme [7] is more practical, where only receiver side CSI (CSIR) is
needed. In MIMO NC, the relay node detects each end node's packet with
traditional MIMO detection and then combines them with network coding. These
schemes failed to exploit the fact that the relay does not need each end
node's individual information. Hence, the performance is limited by
over-detection.

In [8], we have proposed a novel MIMO PNC scheme based on linear detection,
which will be referred as linear MIMO PNC in this paper. Linear MIMO PNC
tries to detect the summation and the difference of the two end node's
packets before transforming them to the network coding form. With similar
complexity and CSIR requirement, it significantly outperforms MIMO NC.
However, due to the performance limit of linear detection, linear MIMO PNC's
performance is poor under bad channel conditions.

Besides the linear  detection, another popular MIMO detection method is
VBLAST (Vertical Bell Laboratories Layered Space-time) which can achieve
much better performance with an acceptable increase in complexity. The
VBLAST architecture was first proposed in [9] where a code block is
de-multiplexed into different layers and each is transmitted through a
particular antenna. At the receiver, these layers are successively detected,
where the detected interference are canceled and the unknown interferences
are nulled by linearly weighting the residual signal vector with a ZF
(zero-forcing) null vector (ZF VBLAST). A low complexity ZF VBLAST scheme is
proposed in [10], where the channel matrix is rewritten in terms of the QR
decomposition as $H=QR$. The inverse of unitary matrix $Q$ was then multiplied to the received
signal before estimating the transmit information. In order to find the
optimal detection order, [10] further proposed the sorted QR decomposition
 algorithm, ZF-SQRD.

In this paper, we combine the basic idea of PNC and the QR VBLAST MIMO
detection scheme and propose VBLAST PNC. Our scheme only requires receiver
side channel state information and symbol level synchronization between the
end nodes, as in general virtual MIMO system. The basic idea of VBLAST PNC
in a 2-by-2 MIMO system is as follows. With QR decomposition, the relay
first detects the second layer (one end node) signal. Rather than canceling
all the component of the second layer signal from the first layer as in
traditional VBLAST detection, we only subtract a part of the second layer
information and directly map the residual signal (including both second and
first layer information) to the network coding form. With such partial
interference cancellation, the error propagated from the incorrectly
detected second layer is significantly decreased. Thus, the system performance is improved.
We then extend our VBLAST
PNC to a detection scheme with an optimal order as in ZF-SQRD, and even
better performance is achieved. Numerical simulation is done to compare the
performance of VBLAST PNC with linear MIMO PNC and MIMO NC schemes. The
results show that VBLAST PNC can achieve much better BER performance than
linear MIMO PNC and VBLAST MIMO NC.


\section{System Model and Illustrating Example}

\begin{figure}[!t]
\centering
\includegraphics[width=3in]{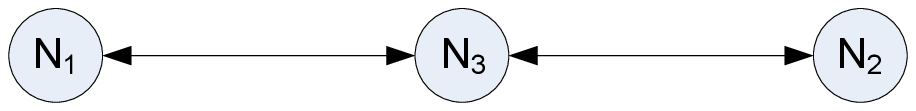}

\caption{Two-Way Relay Channel}
\label{fig_sim}
\end{figure}

\subsection{System Model}
This paper is mainly based on the TWRC in Fig. 1, where the relay is
equipped with 2 antennas and the each end node is equipped with single
antenna.

The transmission consists of two phases. In the uplink phase, both the end
nodes transmit their packets to the relay node simultaneously using QPSK
modulation \footnote{ Our scheme can be easily extended to more general
modulations. }. We assume that the two packets arrive at the relay node in a
symbol level synchronization. In that way, the superimposed signal received
by the relay is:
\begin{equation}
\label{eq1}
\begin{array}{l}
 y_{1} =h_{11} x_{1} +h_{12} x_{2} +n_{1} \\
 y_{2} =h_{21} x_{1} +h_{22} x_{2} +n_{2} \\
 \end{array}
\end{equation}
where $y_{i}$ denotes the received signal at the $i$-th antenna of the relay
node; $h_{i,j}$ is zero mean complex Gaussian random variable, which denotes
the channel coefficient from the end node $N_{j}$ to the $i$-th antenna of the
relay node; $x_{i} \in \{\pm 1\pm \sqrt{-1}\}$is the transmitted signal with
QPSK modulation of the end node$ N_{i}$, and $n_{i}$ demotes the complex
Gaussian noise with zero mean and $\sigma^{2}$ variance for each dimension.

In this phase, the full channel information is available at the relay node.
Rewriting the received signal in the vector form as
\begin{equation}
\label{eq2}
Y=HX+N
\end{equation}
Then, the relay node tries to extract some useful information from $Y$ and
transforming it to the network coded form $x_{1} \oplus x_{2} $. The detailed
processing will be illustrated later, which is also the focus of our paper.

In the downlink phase, the relay broadcasts the network coded packet to both
end nodes. After receiving the packet from the relay, the end nodes extract
their target packets with the help of their own information.

\subsection{Illustrating Example}
Now, we present an example to illustrate the basic idea of VBLAST PNC and its superiority over VBLAST NC,
with a simple channel realization.

In the first phase, the transmission in (2) can be regarded as a $2 \times 2$
MIMO system (two transmit antennas and two received antennas). The relay
node's goal is to acquire an estimate of $x_{1} \oplus x_{2} $. In the
traditional MIMO NC scheme, the relay node decodes $x_{1}$ and $x_{2}$
explicitly before network encoding them. Nevertheless, this scheme is
suboptimal.

Consider an ill-conditioned channel matrix $H$
\begin{equation}
\label{eq3}
H=\left[ {{\begin{array}{*{20}c}
 1 \hfill & {1+\Delta } \hfill \\
 0 \hfill & \Delta \hfill \\
\end{array} }} \right]
\end{equation}

where $\Delta $ is a small quantity. Then eq.(1) can be rewritten as
\begin{equation}
\label{eq4}
\begin{array}{l}
 y_{1} =x_{1} +(1+\Delta )x_{2} +n_{1} \\
 y_{2} =\Delta x_{2} +n_{2} \\
 \end{array}
\end{equation}

With QR ZF VBLAST detection, $x_{2}$ is first detected as
\begin{equation}
\label{eq5}
\widetilde{x_{2} }=\frac{1}{\Delta }y_{2} =x_{2} +\frac{1}{\Delta }n_{2}
\end{equation}
According to the processing of VBLAST, we cancel the interference $x_{2}$ in
$y_{1}$ and obtain the estimate of $x_{1}$ as
\begin{equation}
\label{eq6}
\begin{array}{l}
 \widetilde{x_{1} }=y_{1} -(1+\Delta )\widetilde{x_{2} }\\
 =x_{1} +(1+\Delta)x_{2} +n_{1} -(1+\Delta )(x_{2} +\frac{1}{\Delta }n_{2} ) \\
 =x_{1} +n_{1} -\frac{1+\Delta }{\Delta }n_{2} \\
 \end{array}
\end{equation}

From (\ref{eq5}) and (\ref{eq6}), we can find that the SNR of $x_{1}$ tends to
zero when $\Delta $ is very small. it means that the VBLAST NC scheme
can't accurately estimate the target signal $x_{1} \oplus x_{2} $ in this
case.

Based on the basic idea of PNC, the relay can estimate the target signal
$x_{1} \oplus x_{2} $ from $x_{1} +x_{2} $. Therefore, we can first estimate
$x_{1} +x_{2} $ rather than estimate individual information of $x_{1} ,x_{2}
$, from $Y$. The particular processing is as follows. After estimating $x_{2}$,
we subtract $\Delta \widetilde{x_{2} }$ rather than $(1+\Delta
)\widetilde{x_{2} }$ from $ y_{1}$. Then, we can obtain

\begin{equation}
\label{eq7}
y_{1} '=y_{1} -\Delta \widetilde{x_{2} }\mbox{=}x_{1} +x_{2} +n_{1} -n_{2}
\end{equation}

With the PNC mapping [3], we can directly map $y_{1}'$ in (7) to the target
information  $x_{1} \oplus x_{2} $. Since the noise in (7) is small
(independent of $\Delta )$, the performance of this scheme is much better
than the MIMO NC scheme.

This example indicates that  MIMO PNC may significantly outperform
the traditional MIMO NC. In the following sections, we introduce the
proposed VBLAST PNC scheme in detail.

\section{VBLAST PNC Scheme}
In this section, we first briefly review the MIMO NC scheme based on QR VBLAST detection algorithm [9, 10] for a comparison. After that, we elaborate the proposed VBLAST PNC.

\subsection{MIMO NC based on QR VBLAST}

Consider the 2-by-2 MIMO system in (2). The channel matrix can be decomposed
with QR decomposition so that $H=$\textit{QR}, where the matrix $Q$ is a unitary matrix
(orthogonal columns with unit norm) and $R$ is an upper triangular matrix.
Multiplying the received signal $Y$ by $Q^{H}$, we can obtain the calculated
signal
\begin{equation}
\label{eq8}
W=Q^{H}Y=RX+U
\end{equation}
where $(\cdot )^{H}$ denotes the matrix conjugate transpose and the new noise vector $U=Q^{H}N$
has the same distribution as $N$. The scalar form of $W$ is
\begin{equation}
\label{eq9}
\begin{array}{l}
 w_{1} =r_{11} x_{1} +r_{12} x_{2} +u_{1} \\
 w_{2} =r_{22} x_{2} +u_{2} \\
 \end{array}
\end{equation}
Owing to the upper triangular characteristic of $R$, we can easily estimate the
second layer signal ($x_{2}$ here) as:

\begin{equation}
\label{eq10}
\widetilde{x_{2} }=\frac{w_{2} }{r_{22} }=x_{2} +\frac{u_{2} }{r_{22} }
\end{equation}
Since the QPSK modulation is adopted, the above signal can be demodulated
with hard decision as
\begin{equation}
\label{eq11}
\widehat{x_{2} }=sign(\widetilde{x_{2} })
\end{equation}
Note that the hard decision in (\ref{eq11}) is performed for the real part and the
imaginary part respectively. According to the basic idea of VBLAST, after
the second layer signal $x_{2}$ is detected, we can then detect the first
layer signal ($x_{1}$ here) by canceling the interference of $x_{2\, }$as:
\begin{equation}
\label{eq12}
\widehat{x_{1} }=sign(\widetilde{x_{1} })=sign\left( {\frac{w_{1} -r_{12}
\widetilde{x_{2} }}{r_{11} }} \right)
\end{equation}

After obtaining the individual decisions of $x_{1}$ and $x_{2}$, a
straightforward way to calculating $x_{1} \oplus x_{2} $ is to combine the
estimates of $x_{1}$ and $x_{2}$ (obtained in (11) and (12)
respectively):
\begin{equation}
\label{eq13}
\widehat{x_{1} \oplus x_{2} }=\widehat{x_{1} }\oplus \widehat{x_{2} }
\end{equation}
Hereafter, we refer to this scheme as VBLAST NC. VBLAST NC detects the
signals separately with VBLAST algorithm, and then encodes them into the
network-coded form.However, it may perform poorly as shown in our
illustrating example.

As in the VBLAST processing above,  exchange of the detection order between x1 and x2 (first
detect $x_{1}$ and then cancel it before detecting $x_{2})$ is also workable. As shown in [10],
the detection sequence is crucial to the performance of VBLAST because of
the risk of error propagation. To obtain a better performance, we can
permute the columns of the channel matrix $H$ before the QR decomposition. By
carefully selecting the permutation pattern, the post-permuted $R$ has a larger
$r_{22}$. As a result, the SNR of the second layer is increased and the BER
is decreased, and the error propagation effect is also decreased. We can
also adopt this sorted algorithm [10] in our VBLAST PNC, as elaborated in the following part.

\subsection{VBLAST PNC Scheme}

In VBLAST, the diversity order of the first layer should be larger than the
second layer in theory under the assumption of clean interference
cancellation. In practice, however, the diversity order of both layers is
the same [11]. The main reason is the error propagated from the erroneously
detected second layer signal. To mitigate the error propagation and better
the performance, we propose VBLAST PNC scheme which only cancels a part of
the detected second layer information. After the cancellation, we require
the remaining signal to be in the form of
$x_{1}+{kx}_{2}+u_{1}$ where $k$ is an integer. We can then directly map this
signal to the target signal $x_{1} \oplus x_{2}$ by applying PNC mapping,
without explicitly detecting $x_{1}$. Since the PNC mapping has similar
performances as the ordinary point-to-point transmission, our VBLAST PNC
could achieve a better performance by mitigating the error propagation
effects. The detail of VBLAST PNC is as follows.

We  rewrite $w_{1}$ in (9) as
\begin{equation}
\label{eq14}
w_{1} =r_{11} (x_{1} +kx_{2} )+(r_{12} -kr_{11} )x_{2} +u_{1}
\end{equation}
where $k$ is an integer to be determined later. In (\ref{eq1}) we regard
$x_{1}+$\textit{kx}$_{2}$ as the signal to be estimated and
($r_{12}$\textit{-kr}$_{11})x_{2}$ as interference to be cancelled.

In order to decrease the effect of error propagation, we must minimize
($r_{12}$\textit{-kr}$_{11})$. For example, if $k=r_{12}/r_{11}$, no interference needs
to be cancelled and there will never exist error propagation during the
detection of the signal in (\ref{eq14}). Taking the integer requirement of $k$ into
account, we can determine the value of $k$ as
\begin{equation}
\label{eq15}
k=\lceil {real({r_{12} )} / {r_{11} }} \rfloor
\end{equation}

where $\lceil t \rfloor $ means the integer nearest to $t$. In
(15), $r_{12}$ is a complex variable and $r_{11}$ is a real variable, and we
only take account the real part to calculate $k$ in this paper.

After cancelling the interference with the hard estimate of $x_{2}$ in (11),
we can obtain the estimate of $x_{1}+kx_{2}$ as
\begin{equation}
\label{eq16}
\widetilde{x_{1}+kx_{2} }=\frac{w_{1}-(r_{12} -kr_{11} )\widehat{x_{2}
}}{r_{11} }
\end{equation}
Finally, the relay node estimates $x_{1} \oplus x_{2} $ only from
$x_{1}+kx_{2}$ as long as $k$ does not equal to 0. When k\textgreater
0, intuitively, $abs(x_{1} +kx_{2})$ is larger for $x_{1} \oplus x_{2}
=0$, while it is smaller for $x_{1} \oplus x_{2} =1$. When k\textless 0,
the situation reverse. Then, for each dimension (real part or imaginary
part) signal, the corresponding decision rule is
\begin{equation}
\label{eq17}
\widehat{x_{1} \oplus x_{2} }=sign(k)sign(abs(x_{1} +kx_{2} )-\gamma )
\end{equation}
where $\gamma $ is the decision threshold and its optimal value can be
calculated as in [3]. In high SNR, we can simplify the calculation of
$\gamma $ and set it to $abs(k)$. Then we have
\begin{equation}
\label{eq18}
\widehat{x_{1} \oplus x_{2} }=\left\{ {{\begin{array}{*{20}c}
 {sign(abs(x_{1} +kx_{2} )-k)} \hfill & {\mbox{when\, }k>0} \hfill \\
 {sign(-k-abs(x_{1} +kx_{2} ))} \hfill & {\mbox{when\, }k<0} \hfill \\
\end{array} }} \right.
\end{equation}

To further improve the performance, we can extend our scheme to the sorted VBLAST PNC, where the optimal detection order is chosen as  the tradition VBLAST. In particular, we exchange the columns of $H$ to obtain a
larger $r_{22}$. Then, the VBLAST algorithm is performed on the new $H$.

\section{Numerical Simulation}
In this section, we present the simulation results for the proposed VBLAST
PNC. To compare its performance, we also show the simulation performance of
linear MIMO PNC, the performance of linear MIMO NC and VBLAST NC.

The simulation setting is mainly based on the system model in section II. In
particular, we use QPSK modulation and set the packet length to $10^{6}$.
The wireless channels are assumed to be block fading with each entry of the
channel matrix to be independently complex Gaussian distributed over $N(0, 1)$.
The noise is Gaussian distribution with $N(0, \sigma^{2})$ and the SNR of
the system is defined as $1/\sigma^{2}$. Simulation results are measured in
terms of bit error rate (BER) of $x_{1} \oplus x_{2} $ at the relay node
since the broadcast phase is straightforward.

In Figure 2, we plot the BER performance of different schemes. As shown in
the figure, the proposed MIMO PNC schemes always outperform their
counterparts. Specifically, our VBLAST PNC outperforms VBLAST NC by about
0.5 dB at a BER of $10^{-3}$; the sorted VBLAST PNC outperforms sorted
VBLAST NC by about 1dB. In sorted VBLAST, the more performance improvement
mainly comes from the fact that the average value of $r_{11}$ is smaller and
the interference to be cancelled, $(r_{12} -kr_{11} )x_{2} $, is smaller.

\begin{figure}[!t]
\centering
\includegraphics[width=3in]{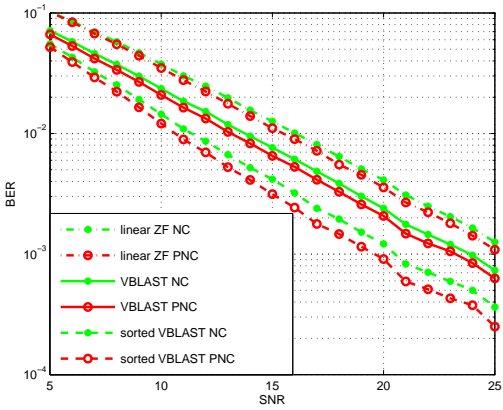}

\caption{BER performance of MIMO PNC and MIMO NC schemes}
\label{fig_sim}
\end{figure}

\section{Conclusion}
In this paper, a novel signal detection and network encoding scheme, VBLAST
PNC, is proposed to extract $x_{1} \oplus x_{2} $ at the relay node in MIMO
TWRC. The basic idea is that the relay node first uses partial interference
cancellation to obtain $x_{1}+$\textit{kx}$_{2\, }$during VBLAST detection process and
then converts it to $x_{1} \oplus x_{2} $ with PNC mapping. With partial
interference cancellation, error propagation effect is mitigated and the
performance is significantly improved. The simulation results verified the
performance advantages of VBLAST PNC under the setting of random Rayleigh
fading channel coefficients. Our scheme is of great interest in practice
since  only CSIR, symbol lever synchronization and low complexity are needed.


%

\appendices


\section*{Acknowledgment}
This work was partially supported by NSFC (No. 60902016), NSF Guangdong (No. 10151806001000003), and NSF Shenzhen (No. JC201005250034A).

\ifCLASSOPTIONcaptionsoff
  \newpage
\fi

\end{document}